\documentclass[twocolumn,showpacs,preprintnumbers,amsmath,amssymb]{revtex4}

\usepackage{graphicx}
\usepackage{dcolumn}
\usepackage{bm}

\begin{document}

\preprint{APS/123-QED}

\title{Spinon, soliton and breather in the spin-1/2 antiferromagnetic chain KCuGaF$_6$}

\author{Izumi Umegaki,$^{1}$ Hidekazu Tanaka,$^{1}$ Nobuyuki Kurita,$^{1}$ Toshio Ono,$^{2}$ Mark Laver,$^{3,\,4}$ Christof Niedermayer,$^{3}$\\
Christian R\"uegg,$^{3}$ Seiko Ohira-Kawamura,$^{5}$ Kenji Nakajima,$^{5}$ and Kazuhisa Kakurai$^{6}$}

\affiliation{$^1$Department of Physics, Tokyo Institute of Technology, Meguro-ku, Tokyo 152-8551, Japan\\
$^2$Department of Physical Science, Osaka Prefecture University, Sakai, Osaka 599-8531, Japan \\
$^3$Laboratory for Neutron Scattering, Paul Scherrer Institute, CH-5232 Villigen PSI, Switzerland\\
$^4$School of Metallurgy and Materials, University of Birmingham, Birmingham, United Kingdom\\
$^5$Materials and Life Science Division, J-PARC Center, Tokai, Ibaraki 319-1195, Japan\\
$^6$Quantum Beam Science Directorate, Japan Atomic Energy Agency, Tokai, Ibaraki 319-1195, Japan}

\date{\today}

\begin{abstract}
Elementary excitations of the $S\,{=}\,1/2$ one-dimensional antiferromagnet KCuGaF$_6$ were investigated by inelastic neutron scattering in zero and finite magnetic fields perpendicular to the $(1, 1, 0)$ plane combined with specific heat measurements. KCuGaF$_6$ exhibits no long-range magnetic ordering down to 50 mK despite the large exchange interaction $J/k_{\rm B}\,{=}\,103$ K. At zero magnetic field, well-defined spinon excitations were observed. The energy of the des Cloizeaux and Pearson mode of the spinon excitations is somewhat larger than that calculated with the above exchange constant. This discrepancy is mostly ascribed to the effective XY anisotropy arising from the large Dzyaloshinsky$-$Moriya interaction with an alternating $\bm D$ vector. KCuGaF$_6$ in a magnetic field is represented by the quantum sine-Gordon model, for which low-energy elementary excitations are composed of solitons, antisolitons and their bound states called breathers. Unlike the theoretical prediction, it was found that the energy of a soliton is smaller than that of the first breather, although the energy of the first breather coincides with that observed in a previous ESR measurement. 
\end{abstract}

\maketitle


\section{Introduction}
Magnetic excitations in most conventional three-dimensional (3D) magnets are well described by the linear spin wave theory (LSWT)~\cite{Anderson,Kubo}, whereas for 1D quantum magnets, the magnetic excitations are qualitatively different from the linear spin wave results because of the strong quantum fluctuation characteristic of one dimension. For the spin-1/2 uniform antiferromagnetic Heisenberg chain described as ${\cal H}\,{=}\,\sum_iJ{\bm S}_i\,{\cdot}\,{\bm S}_{i+1}$, the elementary excitation is domain-wall-like $S^z\,{=}\,1/2$ excitation called a spinon~\cite{dCP,Faddeev}. Its dispersion relation is exactly given by des Cloizeaux and Pearson (dCP)\,\cite{dCP} as
\begin{eqnarray}
{\hbar}{\omega}(q)=\frac{\pi}{2}J\left|\sin{q}\right|,
\label{eq:dCP}
\end{eqnarray}
which is larger than that calculated using LSWT by a factor of ${\pi}/2$. The dCP mode was verified by neutron inelastic scattering experiments~\cite{Endoh,Nagler}. Because neutron scattering observes $S_{\rm tot}^z\,{=}\,0$ and $\pm1$ excitations composed of two spinons, the excitation spectrum has a continuum. The upper bound of the two-spinon continuum is given by~\cite{Yamada,Muller}
\begin{eqnarray}
{\hbar}{\omega}(q)={\pi}J\left|\sin{\frac{q}{2}}\right|.
\label{eq:upper_bound1}
\end{eqnarray}
The dCP mode is observed as the lower bound of the two-spinon continuum.

Some 1D quantum magnets show exotic excited bound states with hierarchical structures when subjected to magnetic fields~\cite{Dender,Oshikawa1,Affleck,Kenzelmann1,Kenzelmann2,Zamolodchikov,Coldea}.
An intriguing system is the spin-1/2 antiferromagnetic chain with a staggered Zeeman term in addition to a uniform Zeeman term. The model is described as
\begin{eqnarray}
\mathcal{H}=\sum_{i} \left[J\bm S_i\cdot \bm S_{i+1}-g{\mu}_{\rm B}HS_i^z-(-1)^ig{\mu}_{\rm B}hS_i^x\right], 
\label{eq:model}
\end{eqnarray}
where $h$ is the staggered field induced by the external field $H$ and is perpendicular to $H$~\cite{Oshikawa1,Affleck,Nagata}. In real materials such as Cu(C$_6$H$_5$COO)$_2$\,$\cdot$\,3H$_2$O (copper benzoate)~\cite{Dender,Nagata}, the staggered field arises from the alternating $\bm g$ tensor and the antisymmetric interaction of the Dzyaloshinsky$-$Moriya (DM) type with the alternating $\bm D$ vector~\cite{Moriya}. 

When the staggered field $h$ is absent, gapless excitations occur in an external magnetic field at the incommensurate wave vectors $q\,{=}\,{\pm}\,2{\pi}m(H)$ and ${\pi}\,{\pm}\,2{\pi}m(H)$ in addition to $q\,{=}\,0$ and ${\pi}$, where $m(H)$ is the magnetization per site in the unit of $g{\mu}_{\rm B}$~\cite{Pytte,Ishimura,Heilmann}.
Because of the staggered field $h$, these gapless excitations have finite gaps. We put $q_0\,{=}\,2{\pi}m(H)$ hereafter. 

Using the field theoretical approach~\cite{Oshikawa1,Affleck}, the low-temperature and low-energy properties of model (\ref{eq:model}) can be represented by the quantum sine-Gordon (SG) model with the Lagrangian density 
\begin{eqnarray}
\mathcal{L}=\frac{1}{2}\left[\left(\frac{{\partial}{\phi}}{{\partial}{t}}\right)^2\,{-}\,v_{\rm s}^2\hspace{-0.5mm}\left(\frac{{\partial}{\phi}}{{\partial}{x}}\right)^2\right]+hC\cos (2{\pi}R{\tilde \phi}),
\label{eq:Lag}
\end{eqnarray}
where $\phi$ is a canonical Bose field, $\tilde \phi$ is the dual field, $R$ is the compactification radius,  $v_{\rm s}$ is the spin velocity and $C$ is a coupling constant. The dual field $\tilde \phi$ corresponds to the angle between the transverse component of the spin and the reference direction in a plane perpendicular to the external magnetic field. Low-energy elementary excitations of the quantum SG model are composed of solitons, antisolitons and their bound states called breathers. The solitons and antisolitons correspond to the excitations at $q\,{=}\,{\pm}\,q_0$ or ${\pi}\,{\pm}\,q_0$. The classical pictures of the solitons and antisolitons are the local rotation and inverse rotation of the spin in a plane perpendicular to the magnetic field, respectively. Their energy $M_{\rm s}$ (soliton mass) is expressed as
\begin{eqnarray}
M_{\rm s}=\frac{2v_{\rm s}}{\sqrt{\pi}}\frac{\Gamma \left(\displaystyle\frac{\xi}{2}\right)}{\Gamma \left(\displaystyle\frac{1+\xi}{2}\right)}\left[\frac{\Gamma \left(\displaystyle\frac{1}{1+\xi}\right)}{\Gamma \left(\displaystyle\frac{\xi}{1+\xi}\right)} \frac{c{\pi}g{\mu}_{\rm B}H}{2v_{\rm s}}c_{\rm s}\right]^{(1+\xi)/2}\hspace{-8mm},\ \ 
\label{eq:solitonmass}
\end{eqnarray}
where $\xi$ is a parameter given by ${\xi}\,{=}\,[2/({\pi}R^2)-1]^{-1}$, $c$ is a parameter depending on magnetic field and $c_{\rm s}\,{=}\,h/H$~\cite{Essler1}. The field dependences of these parameters are shown in the literature~\cite{Affleck,Essler1,Hikihara}. For $H \rightarrow$ 0, $v_{\rm s}\,{\rightarrow}\,({\pi}/2)J$, $\xi\,{\rightarrow}\,1/3$ and $c\,{\rightarrow}\,1/2$. The soliton mass expressed by eq.~(\ref{eq:solitonmass}) is applicable in the wide field range below the saturation field $H_{\rm s}\,{=}\,2J/g{\mu}_{\rm B}$~\cite{Essler1}. The breathers correspond to the excitations at $q\,{=}\,0$ and ${\pi}$ and have hierarchical structures. The mass of the $n$-th breather is determined by the soliton mass $M_{\rm s}$ and the parameter $\xi$ as
\begin{eqnarray}
M_n=2M_{\rm s} {\sin}\,\left(\frac{n{\pi}{\xi}}{2}\right).
\label{eq:breather}
\end{eqnarray}
The number of breathers is limited by $n\,{\leq}\,[{\xi}^{-1}]$\,\cite{Affleck}. In our experimental field range, $g{\mu}_{\rm B}H/J\,{<}\,0.2$, breathers up to the third order can be observed. Because ${\xi}\,{<}\,1/3$ for $H\,{\neq}\,0$~\cite{Affleck,Essler1,Hikihara}, the condition $M_1\,{<}\,M_{\rm s}$ usually holds. Because ${\xi}$ is close to 1/3 at a low magnetic field, the magnetic-field-induced gap ${\Delta}$ is proportional to $H^{2/3}$, which is different from the LSWT result ${\Delta}\,{\propto}\,H^{1/2}$~\cite{Oshikawa1,Affleck,Essler1}. This field dependence of the gap coincides with the experimental gap observed in copper benzoate~\cite{Dender,Asano,Nojiri}. Besides copper benzoate, CuCl$_2$\,${\cdot}$\,2((CD$_3$)$_2$SO)~\cite{Kenzelmann1,Kenzelmann2}, Yb$_4$As$_3$~\cite{Oshikawa2,Kohgi}, and PM\,$\cdot$\,Cu(NO$_3$)$_2\,\cdot$\,(H$_2$O)$_2$ (PM=pyrimidine)~\cite{Feyerherm,Zvyagin,Wolter1,Wolter2} are described by the quantum SG model in external magnetic fields. In these compounds, the exchange constant $J$ ranges from 16 to 36 K and the proportionality coefficient $c_{\rm}\,{=}\,h/H$ is $c_{\rm s}\,{\leq\,}0.08$~\cite{Nojiri,Zvyagin,Kenzelmann2}.

\begin{figure}[t]
	\begin{center}
		\includegraphics[scale =0.52]{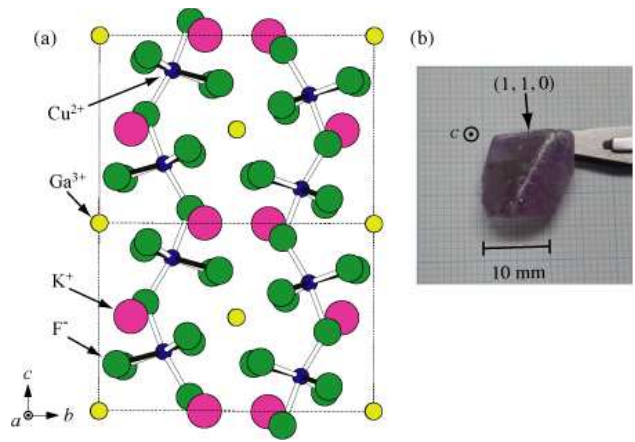}
	\end{center}
	\caption{(Color online) (a) Crystal structure of KCuGaF$_6$ viewed along the $a$ axis. Filled and unfilled bonds respectively denote the elongated and compressed axes of CuF$_6$ octahedra. Dotted lines denote the chemical unit cell. (b) Photograph of KCuGaF$_6$ single crystal used in neutron scattering experiments.}
	\label{fig:KCuGaF6}
\end{figure}

In this paper, we report on the magnetic excitations of KCuGaF$_6$ investigated by inelastic neutron scattering in zero and magnetic fields. KCuGaF$_6$ crystallizes in a monoclinic structure of space group $P2_1/c$~\cite{Dahlke}. The lattice parameters at room temperature are $a\,{=}\,7.2856$\,\AA, $b\,{=}\,9.8951$\,\AA, $c\,{=}\,6.7627$\,{\AA} and $\beta\,{=}\,93.12^\circ$. KCuGaF$_6$ is magnetically described as an $S\,{=}\,1/2$ Heisenberg-like antiferromagnetic chain with the exchange constant of $J/k_{\rm B}\,{=}\,103\,{\pm}\,2$\,K (\,${=}\,8.87\,{\pm}\,0.17\,$\,meV)~\cite{Morisaki,Umegaki1,Umegaki2}. The magnetic chain is running along the crystallographic $c$ axis. Because the CuF$_6$ octahedra are elongated perpendicular to the chain direction owing to the Jahn-Teller effect and the elongated and compressed principal axes of the octahedra alternate along the chain direction, as shown in Fig.~\ref{fig:KCuGaF6}(a), the ${\bm g}$ tensor has the staggered component ${\bm g}_{\rm s}$ along the $c$ direction. Thus, the ${\bm g}$ tensor at the $i$-th site is written as $\bm{g}_{i}\,{=}\,\bm{g}_{\rm u}+(-1)^{i}\bm{g}_{\rm s}$, where $\bm{g}_{\rm u}$ is the uniform ${\bm g}$ tensor~\cite{Umegaki1}. In KCuGaF$_6$, the DM interaction of the form ${\bm D}_i\,{\cdot}\,[{\bm S}_i\,{\times}\,{\bm S}_{i+1}]$ is also allowed between neighboring spins along the chain direction. Consequently, the magnetic model of KCuGaF$_6$ in the external magnetic field $\bm H$ should be expressed as
\begin{eqnarray}
\mathcal{H}=\sum_{i}\left\{J\bm{S}_i{\cdot}\bm{S}_{i+1}-\mu_{\mathrm{B}}(\bm{g}_{i}\bm{H}){\cdot}\bm{S_i}+\bm{D}_{i}{\cdot}[\bm{S}_{i}{\times}\bm{S}_{i+1}]\right\}.\hspace{1mm}
\label{eq:model2}
\end{eqnarray}
Because of crystal symmetry, the ${\bm D}_i$ vector is expressed as ${\bm D}_i=\left[(-1)^iD_x, D_y, (-1)^iD_z\right]$, where the $x$, $y$ and $z$ axes are chosen to be parallel to the $a^*$, $b$ and $c$ axes, respectively~\cite{Umegaki1}. The $a^*$ and $c$ components of the ${\bm D}_i$ vector alternate along the chain direction. If the $b$ component $D_y$ is negligible, then the $\bm D_i$ vector is expressed as ${\bm D}_i\,{=}\,(-1)^i{\bm D}$. 

Owing to the staggered ${\bm g}$ tensor and $\bm D$ vector, the staggered transverse magnetic field ${\bm h}$ is induced perpendicular to the external magnetic field $\bm H$. The staggered field ${\bm h}_i$ acting on the $i$-th site is approximated as~\cite{Oshikawa1,Affleck}
\begin{eqnarray}
\bm{h}_i\simeq(-1)^i\left[\frac{1}{2J}{\bm{H}}\,{\times}\,{\bm D} +\frac{\bm{g}_s}{g}\bm{H}\right].
\label{eq:h_st}
\end{eqnarray}
The magnitude of $\bm{h}_i$ is proportional to $H$. For these reasons, the effective model of KCuGaF$_6$ in an external magnetic field can be written as eq.\,(\ref{eq:model})~\cite{Morisaki,Umegaki1}.

In previous electron spin resonance (ESR) measurements on KCuGaF$_6$, we observed as many as about ten resonance modes for four different field directions at $T\,{=}\,0.5$ K and found that most of the ESR modes can be very well explained by the quantum SG field theory with only the adjustable parameter $c_{\rm s}$~\cite{Umegaki1}. Later, unknown ESR modes that cannot be assigned in the standard quantum SG field theory were successfully explained in terms of the boundary bound states of solitons and breathers~\cite{Furuya}. KCuGaF$_6$ differs from other quantum SG substances in its large exchange interaction of $J/k_{\rm B}\,{=}\,103$\,K and its wide range of the proportionality coefficient $c_{\rm s}\,{=}\,0.031\,{-}\,0.178$~\cite{Umegaki1}. The largest proportionality coefficient in KCuGaF$_6$ is twice as large as those in other quantum SG substances. This indicates that the staggered components of the ${\bm g}$ tensor and DM interaction are considerably large. Thus, KCuGaF$_6$ seems useful for a comprehensive clarification of the elementary excitations of the systems described by model (\ref{eq:model}).

Although ESR is a powerful tool for detecting excitations with high resolution, the detectable excitations are limited to the $q\,{=}\,0$ excitations. To observe the magnetic excitations in a wide $q$ range in zero magnetic field including the dCP mode, spinon continuum and fine structures of the elementary excitations at around $q\,{=}\,{\pi}$ in magnetic fields, we performed inelastic neutron scattering on KCuGaF$_6$. We also measured specific heat down to 50 mK at zero magnetic field to investigate whether magnetic ordering exists. This paper is organized as follows. The experimental procedure is described in Sec. II. Experimental results and discussion are given in Sec. III. Section IV is devoted to conclusions.


\section{Experimental Details}
Single KCuGaF$_6$ crystals were grown from the melt of an equimolar mixture of KF, CuF$_2$ and GaF$_3$ packed into a Pt tube of 14.6 mm inner diameter and 100 mm length. One end of the Pt tube was welded. The other end was tightly folded with pliers after the materials were dehydrated by heating in vacuum at about 150$^{\circ}$C for three days. The materials were treated in a glove box filled with dry nitrogen, because they are hygroscopic. The Pt tube was placed about 10 cm apart from the center of the horizontal furnace to establish a temperature gradient along the Pt tube. The temperature at the center of the furnace was decreased from 800 to 500$^\circ$C at a rate of 1$^{\circ}$C/h. Transparent light-pink crystals were obtained. A single crystal of $14\,{\times}\,14\,{\times}\,5$ mm$^3$ size, as shown in Fig.~\ref{fig:KCuGaF6}(b), was used for neutron inelastic scattering experiments. KCuGaF$_6$ crystals were cleaved along the $(1,1,0)$ plane. The $c$ axis was determined from the intersection of the $(1,\,1,\,0)$ and $(-1,\,1,\,0)$ planes.

Specific heat was measured down to 50 mK at zero magnetic field using a Physical Property Measurement System (Quantum Design, PPMS) with a dilution refrigerator by the relaxation method. A small sample of 7.55 mg was used.

Magnetic excitations in a wide momentum-energy range at zero magnetic field were measured using the cold-neutron disk chopper spectrometer AMATERAS~\cite{Nakajima} installed in the Materials and Life Science Experimental Facility (MLF) at J-PARC, Japan. The sample was mounted in a cryostat with its $c^{*}$ axis in the horizontal plane. The wave vector ${\bm k}_{\rm i}$ of an incident neutron was set to be perpendicular to the $c^{*}$ axis. The scattering intensities in the plane perpendicular to the $c^{*}$ axis were integrated to obtain the statistics of scattering intensity, assuming good one-dimensionality, which is confirmed by the present specific heat measurement shown below. The sample was cooled to 5.8 K using a closed-circle helium refrigerator. Scattering data were collected with three sets of incident neutron energies, $E_{\rm i}\,{=}\,(42.03, 15.19, 7.741, 4.683)$ meV, (94.72, 23.71, 10.54, 5.931) meV and $(34.94, 13.53, 7.133, 4.395)$ meV. All the data were analyzed using the software suite Utsusemi~\cite{Inamura}.

\begin{figure}[bhtp]
	\begin{center}
		\includegraphics[scale =0.5]{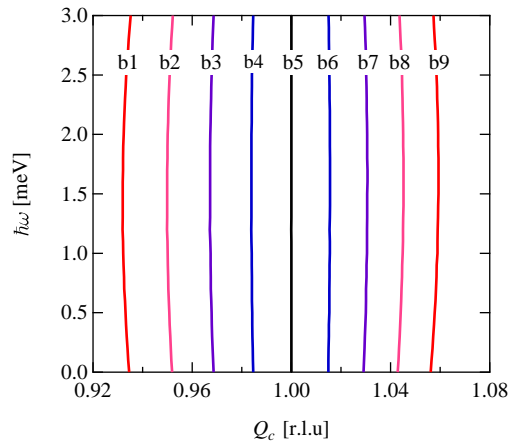}
	\end{center}
	\caption{(Color online) Nine scan loci of the multiblade analyzer in the flat analyzer mode when the center blade (b5) is set to scan at $Q_c\,{=}\,1$ for ${\bm Q}\,{=}\,(0, 0, Q_c)$.}
	\label{fig:blade}
\end{figure}

Low-energy magnetic excitations in magnetic fields were measured using the triple-axis spectrometer RITA-II at SINQ, Paul Scherrer Institute, Switzerland. A multiblade analyzer composed of nine blades is installed on RITA-II. Because the exchange constant of KCuGaF$_6$ is $J\,{=}\,8.87$ meV, the momentum transfer for a soliton is estimated as $Q_{c{\rm s}}^{(\mp)}\,{=}\,0.966$ and 1.034 at $H\,{=}\,10$ T, which is very close to the antiferromagnetic zone center at $Q_c\,{=}\,1$. Therefore, the soliton and the first breather excitations cannot be well separated in the focusing analyzer mode.
In this experiment, we used the flat analyzer mode, for which scattering intensities at nine different wave vectors can be measured at the same time with high $Q$ resolution, although the wave vectors are limited in a small range. The scan loci of individual blades on the momentum-energy plane are shown in Fig.~\ref{fig:blade}. The scan locus for the center blade (b5) is a straight line, while for the other blades, the scan loci are slightly arced. However, within the experimental resolution, the scan loci of individual blades can be approximated as the constant-$Q$ energy scan. The sample was mounted in the cryomagnet with its $(1, 1, 0)$ cleavage plane coincident with the scattering plane. An external magnetic field of up to 12 T was applied perpendicular to the $(1, 1, 0)$ plane. The horizontal collimations were chosen as $40^{\prime}-80^{\prime}-120^{\prime}-40^{\prime}$. The sample was cooled to 40 mK using a dilution refrigerator. The constant-$k_{\rm f}$ mode was adopted with $E_{\rm f}\,{=}\,3.7$ meV. The energy resolution is ${\Delta}E\,{=}\,0.13$ meV. 

\begin{figure}[h]
	\begin{center}
		\includegraphics[scale =0.50]{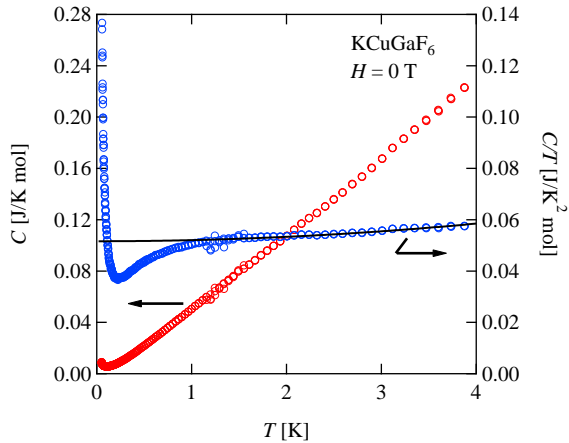}
	\end{center}
	\caption{(Color online) Low-temperature specific heat $C$ and $C/T$ in KCuGaF$_6$ measured at zero magnetic field. The solid line is the fit by $C(T)\,{=}\,{\gamma}T\,{+}\,bT^3$ with ${\gamma}\,{=}\,5.16\,{\times}\,10^{-2}$ J/K$^2$\,mol and $b\,{=}\,4.3\,{\times}\,10^{-4}$ J/K$^{4}$\,mol.}
	\label{fig:heat}
\end{figure}


\section{Results and Discussion}
\subsection{Low-temperature specific heat}

To investigate the spin velocity $v_{\rm s}$, which is the slope of the linear dispersion of spinon excitations at $Q\,{=}\,1$, and the magnetic ordering, we measured specific heat down to 50 mK. Figure~\ref{fig:heat} shows the temperature dependence of the specific heat $C$ and $C/T$ below 4 K measured at zero magnetic field. No magnetic ordering is observed above 50 mK, which indicates that $k_{\rm B}T_{\rm N}/J\,{<}\,5\,{\times}\,10^{-4}$. This result confirms the good one-dimensionality in KCuGaF$_6$ and suggests a ground state of spin liquid. It is noted that $k_{\rm B}T_{\rm N}/J\,{<}\,1\,{\times}\,10^{-3}$ for copper benzoate~\cite{Asano2} and $k_{\rm B}T_{\rm N}/J\,{=}\,2\,{\times}\,10^{-3}$ for Sr$_2$CuO$_3$~\cite{Motoyama}, which are typical spin-1/2 one-dimensional Heisenberg antiferromagnets.

The low-temperature specific heat for an $S\,{=}\,1/2$ antiferromagnetic Heisenberg chain is given by
\begin{eqnarray}
C_{\rm mag}(T)=\frac{{\pi}Rk_{\rm B}}{3v_{\rm s}}T,
\label{eq:heat_LT}
\end{eqnarray}
with $v_{\rm s}\,{=}\,({\pi}/2)J$~\cite{Bonner,Klumper,Takahashi,Johnston}. 
The low-temperature specific heat in KCuGaF$_6$ is approximately proportional to the temperature $T$ down to 1 K, as shown in Fig.~\ref{fig:heat}. The specific heat deviates from the $T$-linear form below 1 K. The upturn in $C$ and $C/T$ below 200 mK arises from the specific heat of nuclear spins. The non-nuclear low-temperature specific heat is described as
$C(T)\,{=}C_{\rm mag}(T)\,{+}\,bT^3$, where the first and second terms denote the magnetic and lattice contributions, respectively. Fitting this equation to $C/T$ for $1.2\,{\leq}\,T\,{\leq}\,4.0$ K, we obtain $v_{\rm s}/k_{\rm B}\,{=}\,169$ K and $b\,{=}\,4.3\,{\times}\,10^{-4}$ J/K$^{4}$\,mol. If the exchange anisotropy is negligible, the exchange constant is evaluated as $J/k_{\rm B}\,{=}\,108\,{\pm}\,1.4$\,K ($\,{=}\,9.31\,{\pm}\,0.12$ meV), which is slightly larger than $J/k_{\rm B}\,{=}\,103\,{\pm}2$\,K as evaluated from magnetic susceptibility data~\cite{Umegaki1}.

\subsection{Magnetic excitations at zero field}

\begin{figure}[b]
  \begin{center}
    \includegraphics[scale =0.50]{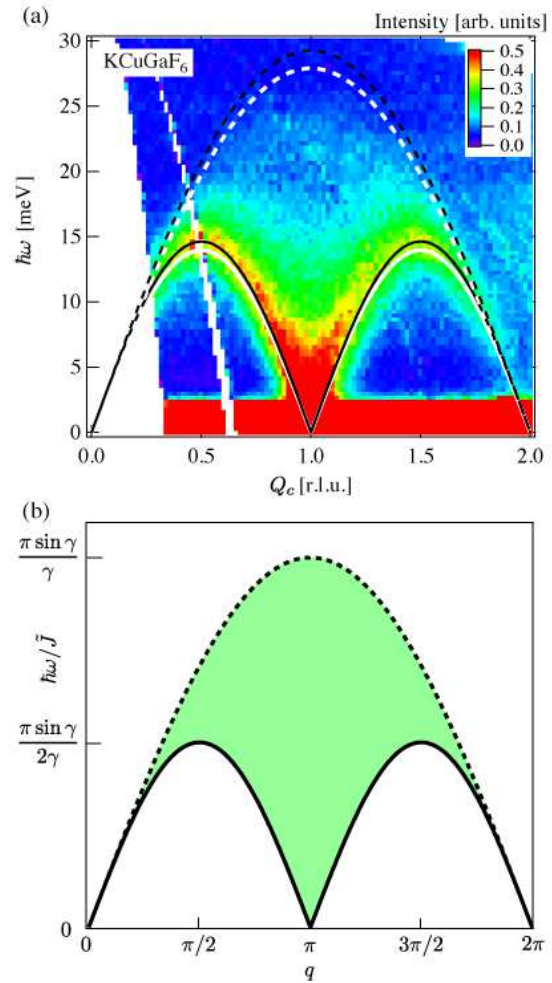}
  \end{center}
  \caption{(Color online) (a) Energy-momentum map of the scattering intensity along the ${\bm Q}\,{=}\,(0, 0, Q_c)$ measured at $T\,{=}\,5.8$ K by using AMATERAS with $E_{\rm i}\,{=}\,34.9$ meV. The white solid and dashed lines express the dCP mode and upper bound of spinon continuum calculated with $J\,{=}\,8.87$ meV, respectively, whereas the black solid and dashed lines represent those calculated with  $J\,{=}\,9.31$ meV. (b) Excitation spectrum of the spin-1/2 XXZ antiferromagnetic spin chain expressed by eq.~(\ref{eq:modelXXZ}) with $0\,\leq\,{\Delta}\,\leq1$.}
  \label{fig:spinon}
\end{figure}

\begin{figure*}[t]
\begin{center}
 \includegraphics[scale =0.50]{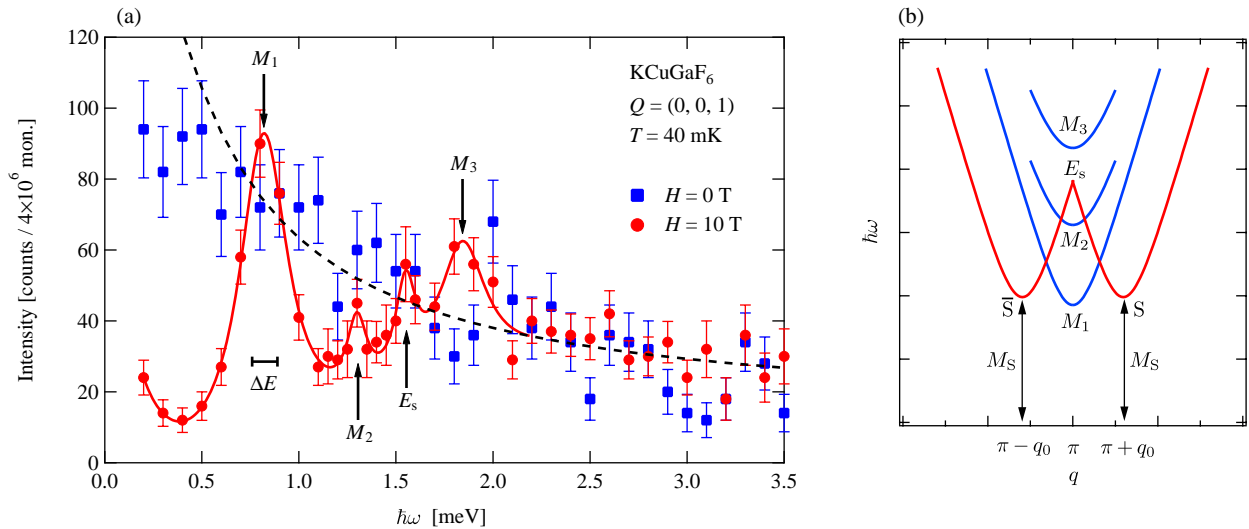}
\end{center}
\caption{(Color online) (a) Energy scans for ${\bm Q}\,{=}\,(0,\,0,\,1)$ at zero magnetic field and 10 T for $H\,{\perp}\,(1,1,0)$ measured at $T\,{=}\,40$ mK. Monitor count $1\,{\times}\,10^6$ approximately corresponds to 7.5 h. Total measuring time at zero magnetic field is half of that at $H\,{=}\,10$ T. Here, the total measuring time is normalized to that at $H\,{=}\,10$ T. The horizontal bar denotes the energy resolution ${\Delta}E$. The dashed line for zero field is a fit of the scattering intensity for a two-spinon cross section using the density of states given by M\"{u}ller et al.\,\cite{Muller}. The solid line is a visual guide assuming four excitations at $H\,{=}\,10$ T. $M_n\ (n\,{=}\,1-3)$ and $E_{\rm s}$ are the excitation energies of the breathers and soliton resonance, respectively, observed in the previous ESR measurement \cite{Umegaki1}. (b) Illustration of the dispersion relations for low-energy excitations at around $q\,{=}\,{\pi}$ expected from the standard quantum SG theory based on model~(\ref{eq:model}). Soliton, antisoliton, soliton resonance and three breathers are labeled as $S$, $\bar S$, $E_{\rm s}$ and $M_1\,{-}\,M_3$, respectively. $M_{\rm s}$ is the soliton mass.} 
\label{fig:excitations_RITAII}
\end{figure*}

Figure~\ref{fig:spinon} shows the energy-momentum map of the scattering intensity along the ${\bm Q}\,{=}\,(0, 0, Q_c)$ obtained at $T\,{=}\,5.8$ K using AMATERAS with the incident neutron energy $E_{\rm i}$ of 34.9 meV. Weak scattering for $10\,{<}\,{\hbar \omega}\,{<}\,20$ meV around $Q_c\,{=}\,2$ arises from the aluminum sample can. The scattering data were collected over 18 hours. The observed scattering spectrum consists of spinon continuum characteristic of the $S\,{=}\,1/2$ Heisenberg antiferromagnetic chain~\cite{Yamada,Muller}. The upper and lower bounds of the spinon continuum have periods of $Q_c\,{=}\,2$ and 1, respectively, because the distance between neighboring Cu$^{2+}$ ions along the chain is $c/2$, as shown in Fig.~\ref{fig:KCuGaF6}(a). The lower bound of magnetic excitations is well described by the dCP mode\,\cite{dCP}. 

The white solid and dashed lines in Fig.~\ref{fig:spinon}(a) are the dCP mode and the upper bound of the spinon continuum calculated using eqs.~(\ref{eq:dCP}) and (\ref{eq:upper_bound1}), respectively, with $J\,{=}\,8.87$ meV, which was determined from the analysis of magnetic susceptibility~\cite{Umegaki1}. The calculated amplitude of the dCP mode is 5\% smaller than the experimental observation. The experimental dCP mode is well described with $J\,{=}\,9.31\,{\pm}\,0.10$ meV, as shown by the black solid line in Fig.~\ref{fig:spinon}(a) within the Heisenberg model. This exchange constant is in good agreement with that obtained from the low-temperature specific heat shown in the previous subsection. 

The exchange constant $J\,{=}\,8.87\,{\pm}\,0.17$ meV, which we used to described the previous ESR results, was evaluated by fitting the theoretical susceptibility characterized by a rounded maximum at $T_{\rm max}\,{=}\,0.641 J/k_{\rm B}$~\cite{Bonner,Eggert,Johnston} to the experimental susceptibility corrected for the Curie term owing to the DM interaction~\cite{Umegaki1}. In the analysis, the error in the absolute value of magnetic susceptibility, which mostly arises from the error in weighing sample mass, is included in the error in the $g$-factor. Therefore, the exchange constant obtained is almost independent of the error in weighing sample mass, and the value of $J$ is fairly accurate. Thus, the increase in excitation energy by $5\,{\pm}\,2$\% from that calculated using $J\,{=}\,8.87$ meV cannot be attributed to the error in the evaluation of the exchange constant $J$. In what follows, we discuss the contribution of the DM interaction to the spinon excitation.

Here, we assume that the $b$ axis component of the ${\bm D}_i$ vector for the DM interaction is negligible in KCuGaF$_6$ and that the $\bm D_i$ vector is expressed as ${\bm D}_i\,{=}\,(-1)^i{\bm D}$. When taking the $z$ axis parallel to $\bm D$, we rotate the local coordinates around the $z$ axis by an angle of ${\pm}\,{\alpha}/2$ for odd and even number spins, respectively, where $\tan {\alpha}\,{=}\,D/J$~\cite{Affleck}. Then the original model (\ref{eq:model2}) at zero magnetic field is transformed into the XXZ model
\begin{eqnarray}
{\cal H}={\tilde J}\sum_i\left(S_i^{x}S_{i+1}^{x}+S_i^{y}S_{i+1}^{y}+{\tilde{\Delta}}S_i^zS_{i+1}^z\right),
\label{eq:modelXXZ}
\end{eqnarray}
where ${\tilde J}\,{=}\,J/{\tilde{\Delta}}$ and ${\tilde{\Delta}}\,{=}\,J/\sqrt{J^2+D^2}$.
For $0\leq {\tilde{\Delta}}\leq 1$, the dispersion relation of the spinon excitation is expressed as \cite{Takahashi}
\begin{eqnarray}
{\hbar}{\omega}(q)=\frac{{\pi}\sin{\gamma}}{2{\gamma}}{\tilde J}\left|\sin{q}\right|,
\label{eq:dCPXXZ}
\end{eqnarray}
with ${\gamma}\,{=}\,\cos^{-1}{\tilde{\Delta}}$. Figure~\ref{fig:spinon}(b) shows the excitation spectrum based on eq.~(\ref{eq:dCPXXZ}). The energy scale is modified by the factor $\sin {\gamma}/{\gamma}$ as compared with the Heisenberg case.
Fitting eq.~(\ref{eq:dCPXXZ}) with $J\,{=}\,8.87\,{\pm}\,0.17$ meV to the experimental excitation spectrum shown in Fig.~\ref{fig:spinon}(a), we obtain $\sin {\gamma}/({\gamma}{\tilde{\Delta}})\,{=}\,1.05\,{\pm}\,0.02$. This leads to a large $\bm D$ vector of $D/J\,{=}\,0.39\,{\pm}\,0.08$. In KCuGaF$_6$, the largest proportionality coefficient $c_{\rm s}\,{=}\,0.178$ is observed for $H\,{\parallel}\,c$~\cite{Umegaki1}. If the staggered field in this case arises only from the DM interaction, the magnitude of the $\bm D$ vector is estimated from eq.~(\ref{eq:h_st}) as $D/J\,{=}\,0.36$. This value of $D$ is consistent with that evaluated from the excitation data within the error range. Thus, we infer that the large DM interaction gives rise to most of the increase in excitation energy. Such a large DM interaction as observed in KCuGaF$_6$ is sometimes observed in fluoride magnets, e.g., $D/J\,{=}\,0.29$ in Cs$_2$Cu$_3$SnF$_{12}$~\cite{Ono} and $D/J\,{=}\,0.18$ in Rb$_2$Cu$_3$SnF$_{12}$~\cite{Matan}.

\subsection{Low-energy excitations in magnetic fields}

Figure~\ref{fig:excitations_RITAII}(a) shows the energy scans for ${\bm Q}\,{=}\,(0,\,0,\,1)$ at zero magnetic field and 10 T for the magnetic field $\bm H$ perpendicular to the $(1,1,0)$ plane measured at $T\,{=}\,40$ mK. For zero magnetic field, the two-spinon excitation starts from zero energy and its intensity decreases monotonically with increasing energy. This is consistent with the intensity map shown in Fig.~\ref{fig:spinon}(a). The dashed line in Fig.~\ref{fig:excitations_RITAII}(a) is a fit using the density of states given by M\"{u}ller {\it et al.}~\cite{Muller}.

\begin{figure}[t]
\begin{center}
\includegraphics[scale =0.50]{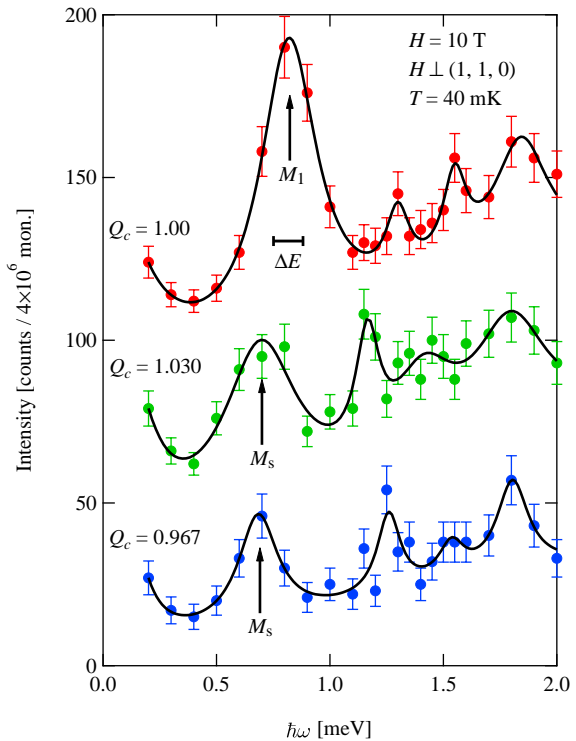}
\end{center}
\caption{(Color online) Energy scans for $Q_c\,{=}\,0.967$, 1.00 and 1.05 at $H\,{=}\,10$ T for $H\,{\perp}\,(1,1,0)$. The intensity data are shifted upward by multiples of 50 counts\,$/\,4\,{\times}\,10^6$ mon. The horizontal bar denotes the energy resolution ${\Delta}E$. $M_1$ and $M_{\rm s}$ are peaks corresponding to the first breather and soliton, respectively. Solid lines are visual guides.} 
\label{fig:soliton_breather}
\end{figure}

\begin{figure}[htb]
\begin{center}
   \includegraphics[scale =0.50]{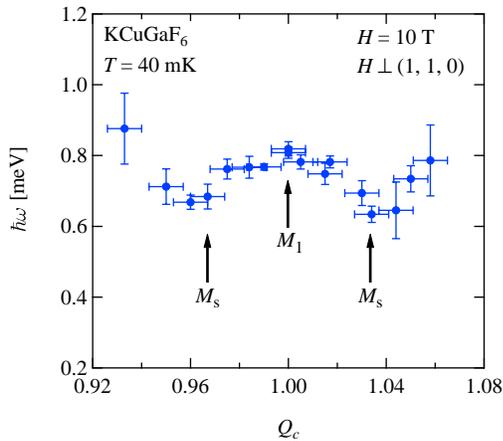}
  \end{center}
   \caption{(Color online) Dispersion relations of the lowest excitations for $H\,{=}\,10$ T. Arrows labeled $M_1$ and $M_{\rm s}$ indicate the wave vectors corresponding to the first breather and soliton, respectively.}
  \label{fig:dispersion}
\end{figure}

At $H\,{=}\,10$ T, the intensity for ${\hbar}{\omega}\,{\leq}\,0.5$ meV decreases markedly to the background level and a strong peak emerges at ${\hbar}{\omega}\,{=}\,0.82$ meV. This result clearly shows that an excitation gap opens in a magnetic field. Figure~\ref{fig:excitations_RITAII}(b) shows the structure of the low-energy excitation at around $q\,{=}\,{\pi}$, which is expected from the standard quantum SG theory based on model~(\ref{eq:model})~\cite{Zvyagin}. In the previous ESR measurements for $H\,{\perp}\,(1, 1, 0)$, we observed three breathers $M_n$ ($n\,{=}\,1-3$) and one soliton resonance $E_{\rm s}$, which is given by $E_{\rm s}\,{\simeq}\,\sqrt{M_{\rm s}^2+(g\mu_{\mathrm{B}}H)^2}$~\cite{Umegaki1}. Their energies at $H\,{=}\,10$ T were obtained as $M_1\,{=}\,0.79$ meV, $M_2\,{=}\,1.32$ meV, $M_3\,{=}\,1.82$ meV and $E_{\rm s}\,{=}\,1.57$ meV. The proportionality coefficient for $H\,{\perp}\,(1, 1, 0)$ was determined as $c_{\rm s}\,{=}\,0.056$. From these ESR results, the strong peak observed at 0.82 meV can be assigned as the first breather $M_1$. In addition to the $M_1$ peak, weak peaks are observed at ${\hbar}{\omega}\,{=}\,1.29$, 1.56 and 1.84 meV. The solid line in Fig.~\ref{fig:excitations_RITAII} is a visual guide assuming four excitations. These weak peaks are considered to be $M_2$, $E_{\rm s}$ and $M_3$, respectively, because their energies are close to those observed by ESR measurements, as indicated by arrows in Fig.~\ref{fig:excitations_RITAII}. However, we do not discuss these weak peaks in this paper, because their statistics are insufficient. We focus on the first breather $M_1$ and the soliton shown below.

The wave vector corresponding to the soliton in KCuGaF$_6$ is estimated as $Q_{c{\rm s}}^{(-)}\,{=}\,0.966$ and $Q_{c{\rm s}}^{(+)}\,{=}\,1.034$ at $H\,{=}\,10$ T for $H\,{\perp}\,(1,1,0)$ using $J\,{=}\,8.87$ meV and $g\,{=}\,2.32$~\cite{Umegaki1}. As shown in Fig.~\ref{fig:blade}, when the center blade (b5) in the analyzer is set to scan at $Q_c\,{=}\,1.00$, the wave vectors for the third and seventh blades (b3 and b7) are $Q_c\,{=}\,0.967$ and 1.030, respectively, which are very close to $Q_{c{\rm s}}^{(\pm)}$. Figure~\ref{fig:soliton_breather} shows the energy scans at $H\,{=}\,10$ T for $Q_c\,{=}\,1.00$, 0.967 and 1.030 using blades b5, b3 and b7, respectively. Solid lines are visual guides. In each scan, one strong peak is observed below 1 meV. The linewidths of these strong peaks are larger than the energy resolution ${\Delta}E\,{=}\,0.13$ meV. Here, we pay attention only to these lowest strong excitations. Excitations corresponding to the soliton are observed at ${\hbar}{\omega}\,{\simeq}\,0.69$ meV indicated by arrows labeled $M_{\rm s}$ for $Q_c\,{=}\,0.967$ and 1.030. Unexpectedly, the soliton mass $M_{\rm s}$ is smaller than the mass of the first breather, $M_1$. Within the standard quantum SG theory based on model~(\ref{eq:model}), the parameter ${\xi}$ in eq.~(\ref{eq:breather}) is smaller than 1/3 in finite magnetic field~\cite{Affleck,Essler1,Hikihara}. Thus, usually, $M_{\rm s}$ is greater than $M_1$ as observed in CuCl$_2$\,${\cdot}$\,2((CD$_3$)$_2$SO)~\cite{Kenzelmann1,Kenzelmann2}.

Figure~\ref{fig:dispersion} shows the dispersion relation of the lowest excitation branch at around $Q_c\,{=}\,1$, which was obtained by using all the blades in the analyzer. The lowest excitation occurs near $Q_c\,{=}\,Q_{c{\rm s}}^{(\pm)}$ corresponding to the wave vectors of solitons. The same behavior of the dispersion relation was also observed for $H\,{=}\,12$ T, in which $Q_{c{\rm s}}^{(+)}\,{=}\,0.959$ and $Q_{c{\rm s}}^{(-)}\,{=}\,1.041$. Theory predicts that the soliton and breather excitations belong to different branches, as shown in Fig.~\ref{fig:excitations_RITAII}(b). However, in KCuGaF$_6$, these two excitations appear to be on the same branch within the experimental resolution, as shown in Fig.~\ref{fig:dispersion}.

\begin{figure}[t]
\begin{center}
   \includegraphics[scale =0.50]{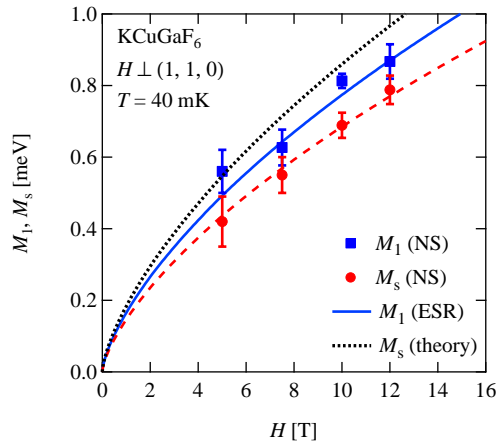}
  \end{center}
   \caption{(Color online) Masses of the first breather, $M_1$, and soliton, $M_{\rm s}$, as a function of magnetic field from our inelastic neutron scattering measurements. The solid line is a fit to $M_1$ obtained by previous ESR measurements using eqs.~(\ref{eq:solitonmass}) and (\ref{eq:breather}) with $J\,{=}\,8.87$ meV, $g\,{=}\,2.32$ and $c_{\rm s}\,{=}\,0.056$~\cite{Umegaki1}. The dotted line is the soliton mass calculated using these parameters. The dashed line is a guide to the eye of the soliton mass observed in this work.}
  \label{fig:mass_SB}
\end{figure}

Figure~\ref{fig:mass_SB} shows $M_1$ and $M_{\rm s}$ measured at four different magnetic fields. The solid line is a fit to $M_1$ obtained by the previous ESR measurements using eqs.~(\ref{eq:solitonmass}) and (\ref{eq:breather}) with $J\,{=}\,8.87$ meV, $g\,{=}\,2.32$ and $c_{\rm s}\,{=}\,0.056$~\cite{Umegaki1}. The dotted line is $M_{\rm s}$ calculated using eq.~(\ref{eq:breather}) with these parameters, which is greater than the calculated $M_1$ by a factor of 1.11. The dashed line is a visual guide of the soliton mass obtained in this work. The mass of the first breather observed by this neutron inelastic scattering experiment is in good agreement with that observed by the ESR measurements, which is approximately in proportion to $H^{2/3}$. However, the experimental soliton mass is smaller than the calculated soliton mass by a factor of 1.25.

In the previous ESR measurements on KCuGaF$_6$~\cite{Umegaki1}, most of the observed modes for $H\,{\perp}\,(1, 1, 0)$ were well described in terms of the standard quantum SG field theory based on model~(\ref{eq:model}) with one adjustable parameter $c_{\rm s}\,{=}\,h/H$. In ESR measurements, only ${\bm Q}\,{=}\,0$ excitations were observed, which are the same as those for ${\bm Q}\,{=}\,(0, 0, 1)$ in KCuGaF$_6$, and thus the soliton excitations cannot be directly observed by ESR measurements. The present neutron scattering experiment reveals that the soliton mass $M_{\rm s}$ is smaller than the mass of the first breather, $M_1$. This result appears to be difficult to understand within model~(\ref{eq:model}). In principle, the magnitude of $M_1$ can become larger up to $2M_{\rm s}$ from eq.~(\ref{eq:breather}) when the parameter $\xi$ is greater than 1/3. This condition is not satisfied unless the effect of the anisotropy is taken into account in calculating the parameter $\xi$. However, the number of breathers, $n$, decreases from three to two if ${\xi}\,{>}\,1/3$ because $n$ is limited by $n\,{\leq}\,[{\xi}^{-1}]$. This is inconsistent with the fact that all the breathers up to the third order were actually observed in the ESR measurements~\cite{Umegaki1}. The effect of the $b$ axis component of the ${\bm D}$ vector, which does not alternate along the chain, also has to be investigated. However, the $b$ axis component is expected to be small, because it is deduced from the Curie terms of the magnetic susceptibilities for three different field directions~\cite{Umegaki1} that the alternating component of ${\bm D}$ vector is approximately oriented to the $a$ axis. A theoretical description of the inversion between the magnitudes of $M_{\rm s}$ and $M_1$ is an ongoing problem.


\section{Conclusions}
In conclusion, we have presented the results of the specific heat and neutron inelastic scattering experiments on the $S\,{=}\,1/2$ 1D antiferromagnet KCuGaF$_6$, which is described by the quantum SG model in a magnetic field. No magnetic ordering is observed above 50 mK despite the large exchange interaction $J\,{=}\,8.87$ meV, which confirms the good one-dimensionality in KCuGaF$_6$. At zero magnetic field, well-defined spinon excitations were observed. The spin velocity obtained from the spinon dispersion coincides with that obtained from the low-temperature specific heat. The observed dCP mode is 5\% greater than that calculated with $J\,{=}\,8.87$ meV. This small disagreement can be explained in terms of the effective XY anisotropy due to the large DM interaction with an alternating $\bm D$ vector. The magnetic excitations near $Q_c\,{=}\,1$ in magnetic fields were investigated in detail using a multiblade analyzer. The massive excitations of the soliton and first breather were clearly observed. However, different from the theoretical prediction, the mass of the first breather is greater than the soliton mass. We suggest that the large magnetic anisotropy is responsible for the inversion between these two excitation energies. \\


\begin{acknowledgments}
We express our sincere thanks to K. Hida and S. Furuya for fruitful discussions and comments. This work was supported by a Grant-in-Aid for Scientific Research (A) (Grant Nos. 23244072 and 26247058) from the Japan Society for the Promotion of Science (JSPS), and by the Global COE Program ``Nanoscience and Quantum Physics'' at Tokyo Tech funded by the Japanese Ministry of Education, Culture, Sports, Science and Technology. I.U. and H.T. were supported by a JSPS Research Fellowship for Young Scientists and a grant from the Mitsubishi Foundation, respectively. 
The experiments on AMATERAS were performed with the approval of J-PARC (Proposals No. 2013B0167). This work is based on experiments performed at the Swiss spallation neutron source SINQ, Paul Scherrer Institute, Villigen, Switzerland. I.U., T.O. and H.T. express their gratitude to the Paul Scherrer Institute for their kind hospitality during the experiment.
\end{acknowledgments}


\end{document}